\documentclass[a4paper]{jpconf}
\usepackage{epsfig}
\usepackage{graphicx}
\usepackage{amsmath}
\usepackage{amssymb}
\usepackage{lipsum}
\usepackage{cite}
\makeatletter
\newcommand{\ket}[1]{| #1 \rangle}
\newcommand{\bra}[1]{\langle #1 |}

\begin{document}

\title{Towards microscopic optical potentials in deformed nuclei}

\author{A Idini, J Rotureau, J Bostr\"om, J Ljungberg, B G Carlsson}

\address{Division of Mathematical Physics, Physics dept., LTH, Lund University, S-22100 Lund, Sweden}

\ead{andrea.idini@matfys.lth.se}

\begin{abstract}
A microscopic and consistent description of both nuclear structure and reactions is instrumental to extend the predictivity of models calculating scattering observables. In particular, this is crucial in the case of exotic nuclei not yet discovered. 

In this manuscript, we will present the plan of the Lund effort for a symmetry breaking description of bound and scattering observables using a generator coordinate method model based on an effective Hamiltonian constrained with a (Skyrme) functional. Following this, we will illustrate the steps to construct an optical potential for deformed nuclei from microscopic wavefunctions obtained with projected generator coordinate method from Hartree-Fock-Bogoliubov basis.
\end{abstract}

\section{Introduction}

It has been proven very difficult to study nuclear reactions using state--of--the--art nuclear structure information in a consistent framework for many nuclei of interest, cf. \cite{Johnson:20, Hebborn:22} and refs. therein for context regarding the importance of a consistent description of nuclear structure and reactions. Despite efforts and a huge progress in later years in developing optical potentials from first principles using Green's function based methods \cite{Idini:18,Idini:19,Rotureau:17,Rotureau:18} or multi--scattering \cite{Holt:13,Vorabbi:21, Burrows:20} approaches, it is especially difficult to develop microscopic potentials for deformed nuclei. The description of deformed nuclei imposes an additional complexity to the calculation and formalism. 

The following is a short summary concerning the construction of a model of atomic nuclei that relies on an effective Hamiltonian and generator coordinate method, with the introduction of the steps to derive optical potentials for deformed nuclei from microscopic symmetry breaking calculation.

\section{Method}

Models of quantum many--body systems try to provide a microscopic description of a system composed of many particles affected by a general Hamiltonian $\hat H$ with the best degree of approximation. It is convenient to notice that a general Hamiltonian in tensor form, can be decomposed in a sum of a product of lower--rank tensors (cf. e.g. \cite{Tichai:19}). Without loss of generality, it is possible to assume these lower rank tensors to be expressed in a multipole expansion, resulting in a general Hamiltonian to be expressed and expanded as product of multipoles in order to exploit the natural degrees of freedom of nuclei.

Following this principle, we developed a formalism and code to provide multi--reference solutions of Hamiltonians expressed in separable multipoles using generator coordinate method (GCM) with symmetry restoration. At the present moment, we focus our attention on a simple Hamiltonian that captures the physical properties of nuclei with the least amount of terms. It was shown using particle--vibration coupling that it is possible to achieve a satisfactory agreement with spectroscopic properties of nuclei using an Hamiltonian composed of separable multipole terms and seniority pairing, with the low--lying quadrupole being the most important contribution \cite{Idini:12,Idini:15}.
\begin{figure}[h!]
\centering
\includegraphics[width=0.9\textwidth]{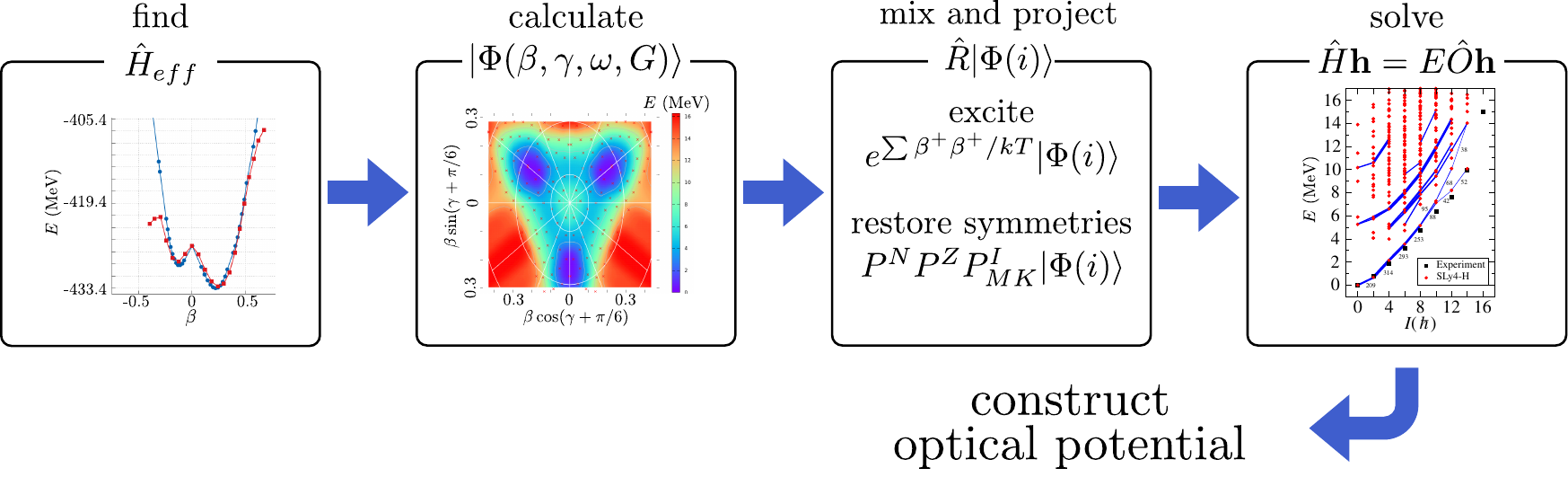} 
\caption{A schematic representation of the method expressed in \cite{Ljungberg:22}, cf. the main text.}
\label{fig:scheme}
\end{figure} 

Therefore, the effective Hamiltonian for \cite{Ljungberg:22} (that was used for the calculation of superheavy isotopes \cite{Samark:21}) is postulated to be, 
\begin{equation}
\hat{H}=\hat{H}_{0}+\hat{H}_{Q}+\hat{H}_{P},\label{eq:H_m}
\end{equation}
that is, $H$ includes a spherical mean field $H_{0}=\sum_{i} e_{i}a_{i}^{\dagger}a_{i} + E_{0}$, and two separable terms for the pairing $H_{P}=-G\sum_{ijkl}P_{ij}P_{kl}a_{i}^{\dagger}a_{j}^{\dagger}a_{k}a_{l}$ and the quadrupole-quadrupole interaction $\hat{H}_{Q}=-\frac{1}{4}\chi\sum_{ijkl}\sum_{\mu}\left[Q_{ij}^{2\mu}Q_{kl}^{2\mu*}-Q_{ik}^{2\mu}Q_{jl}^{2\mu*}\right]a_{i}^{\dagger}a_{l}^{\dagger}a_{k}a_{j}$. The creator operator refers to an element of the spherical basis $i\equiv(n_{i}l_{i}j_{i}m_{i})$, $P_{ij}$ is the seniority pairing matrix element with $G$ the pairing strength fixed according to the uniform spectra method \cite{Ragnarsson:05}, and $Q_{ij}$ the modified quadrupole operator with a radial profile. The Hamiltonian is fitted to reproduce axially constrained calculations of e.g. SLy4 parametrization of Skyrme functional, generating the Sly4-H effective Hamiltonian. More information can be found in \cite{Ljungberg:22}, with comparison of different Skyrme functionals in \cite{Ljungberg:22proc}.

This Hamiltonian is chosen because it is a simple effective Hamiltonian that explicitly breaks particle number and rotational symmetries. In this way, we can describe the many--body properties as comprehensively as possible in a large scale calculation, projecting to good angular momentum $I$ and particle number.

The many-body basis that is used to solve the effective Hamiltonian in Eq. (\ref{eq:H_m}) consists of HFB vacuua $\ket{\Phi}$ with a variation over a set of generator coordinates, which are the familiar $\beta$ and $\gamma$ for deformation and triaxiality, pairing strengths for protons and neutrons $g_{p,n}$, and cranking frequencies $\omega$. Therefore, a set of HFB vacua for different coordinates $\{ \ket{\Phi (\beta,\gamma,g_n, g_p, \omega)} \} \equiv \{ \ket{\Phi(i)} \}$ is obtained. This choice accounts for the most important collective degrees of freedom of collective vibrations, rotations and pairing vibrations.

The GCM solution can be written as,
\begin{equation}
\ket{\Psi^A_J (a)} = \sum_{iK} h^{A,a}_{JMK}(i) P^N P^Z P^{J}_{MK} \ket{\Phi(i)},
\label{eq:GCM_state}
\end{equation}
where the state $a$ of total number of nucleons $A$ is obained summing over the basis states $\ket{\phi(i)}$, considering the coefficients of the solutions of the Hill--Wheeler equation $h_i$, and restoring the symmetry with projection operators $P^N$, $P^Z$, and $P^J_{MK}$ projecting to good number of neutrons $N$, protons $Z$ and good angular momentum $J$. Note that GCM guarantees that the solutions consist in a orthonormal basis with a completeness relation.

The Green's function can, in principle, be constructed with several many-body methods from densities and eigenstates (cf. \cite{Rotureau:17,Idini:19,Salvioni:20}). It is defined from the eigenstates of $H$,
\begin{align}
G(\alpha,\beta,E,\eta) = & \bra{\Psi^A_0} a_\alpha \frac{1}{E-(H-E^A_0)+i\eta} a^\dagger_\beta\ket{\Psi^A_0}
+ \bra{\Psi^A_0} a^\dagger_\beta \frac{1}{E + (H-E^A_0)-i\eta} a_\alpha \ket{\Psi^A_0},
\label{eq:Green}
\end{align}
with $E^A_0$ as the energy of the ground state $\Psi^A_0$ for system of $A$ particles.
In the case under consideration, the reference state is an even--even state $A$, implying that the ground state spin is $0$. After the insertion of the identity resolution $\sum \ket{\Psi^{A+1}_J(k)} \bra{\Psi^{A+1}_J(k)}$ the matrix elements of the creation and annhilation operators define the components of the numerator of the Green's function and are called the spectroscopic factors (SF). Together with the energy of the states, the SF are the elements needed to construct the Green's functions (\ref{eq:Green}), hence the optical potential \cite{Johnson:20}.

The familiar basis associated to the creation operator $a^\dagger$ can be either a spherical Hartree Fock or harmonic oscillator basis. In this case, the spectroscopic factors between GCM states are obtained following (\ref{eq:GCM_state}) considering the overlap between even and odd states,
\begin{equation}
S^J(k,\alpha) = \sum_{iaK} h^{A,0}_{00}(i) (h^{A+1,k}_{JK}(a))^* \bra{\Phi^{A+1}(a)} a^\dagger_{\alpha K} P^N P^Z P^{0}_{00} \ket{\Phi(i)}.
\end{equation}
Therefore, in the Bogoliubov quasiparticle basis $\beta$ the calculation of $S^J(k,\alpha)$ is then composed of two contributions:
$\bra{\Phi^{A}(z)} P^N P^Z P^{0}_{00} \ket{\Phi(i)}$ and $\bra{\Phi^{A}(z)} \beta_q \beta_l P^N P^Z P^{0}_{00} \ket{\Phi(i)}$, cf. \cite{Bostrom:22proc}.

\section{Conclusions}

The provided definition of SF in the Bogoliubov basis for GCM states can be used to calculate them, hence defining Green's functions, optical potentials, and calculate reaction observables consistently.  An important seminal example of use of GCM in a small--scale calculation of SF can be found in \cite{Haakansson:78}. The first results of the calculation of the SF for $^{24}$Mg are in \cite{Bostrom:22proc}, using these results we plan to progress to the construction of optical potentials as described here.

\paragraph*{Acknowledgments}
We were supported by Swedish Research Council 2020-03721 and Crafoord fundation and used the Lunarc computing facility.

\section*{References}


\bibliographystyle{iopart-num}
\bibliography{nuclear}
\end{document}